\documentclass[prb,twocolumn,preprintnumbers,amsmath,amssymb]{revtex4}

\usepackage{amsmath,amssymb,amsmath,latexsym,epsf,epsfig,graphicx}

\pdfoutput=1

\newcommand{\II}{\mbox{${\mathbb I}$}}

\newcommand{\RR}{\mbox{${\mathbb R}$}}

\newcommand{\ZZ}{\mbox{${\mathbb Z}$}}

\def\S{\mathbb S}

\def\tc{\mathbf t}
\def\bA{\mathbf A}
\def\bx{\mathbf x}

\newcommand{\rd}{{\rm d}}

\newcommand{\bea}{\begin{eqnarray}}
\newcommand{\eea}{\end{eqnarray}}

\def\cF{\mathcal F}

\def\ri{{\rm i}}

\def\e{{\rm e}}

\begin{document}

\title{Quantum Wire Network with Magnetic Flux}
\author{Vincent Caudrelier$^1$, Mihail Mintchev$^2$ and Eric Ragoucy$^3$}

\affiliation{
${}^1$ Center for Mathematical Science, City University London, 
Northampton Square, London EC1V 0HB, UK \\
${}^2$ Istituto Nazionale di Fisica Nucleare and Dipartimento di Fisica dell'Universit\`a di Pisa, 
Largo Pontecorvo 3, 56127 Pisa, Italy\\ 
${}^3$ LAPTh, Laboratoire d'Annecy-le-Vieux de Physique Th\'eorique, 
CNRS, Universit\'e de Savoie,   
 BP 110, 74941 Annecy-le-Vieux Cedex, France}

\date{\today}

\begin{abstract} 

The charge transport and the noise of a quantum wire network, 
made of three semi-infinite external leads attached to a ring crossed by a 
magnetic flux, are investigated. The system is driven away from equilibrium
by connecting the external leads to heat reservoirs with different 
temperatures and/or chemical potentials. The properties of the exact scattering 
matrix of this configuration as a function of the momentum, the magnetic flux and the transmission 
along the ring are explored. We derive the conductance and the noise, describing in 
detail the role of the magnetic flux. In the case of weak coupling between the ring and the 
reservoirs, a resonant tunneling effect is observed. We also discover that a non-zero 
magnetic flux has a strong impact on the usual Johnson-Nyquist law for the  
pure thermal noise at small temperatures.

\end{abstract}

\maketitle

\section{Introduction}

\begin{figure}[ht]
\begin{center}
\begin{picture}(220,90)(-20,0) 
\includegraphics[scale=0.8]{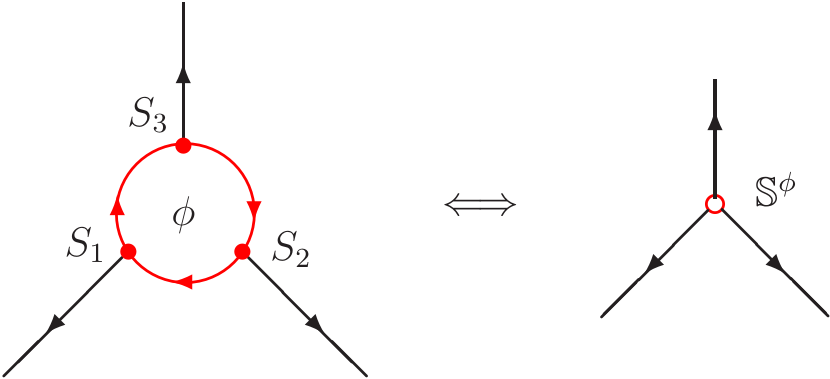}
\end{picture} 
\end{center}
\caption{(Color online) Ring junction with local $S$-matrices $\{S_j\, :\, j=1,2,3\}$ and 
magnetic flux $\phi$ (left) and its Y-junction counterpart with equivalent total scattering 
matrix $\S^\phi $ (right).}
\label{fig.1}
\end{figure}

In this paper we investigate the effect of ambient electromagnetic fields on quantum wire networks. 
We focus on the network displayed in left hand side of fig.~\ref{fig.1}, composed of  
a ring enclosing a magnetic flux $\phi$, and three semi-infinite leads. 
The interactions at the vertices $V_j$ are described by local scattering matrices $S_j$, whereas 
along both the internal and external edges $E_i$ the charges 
interact with a time-independent ambient electromagnetic field, 
generated by a classical potential $\bA(\bx)$. We show that all these interactions can be incorporated in 
a {\it fully equivalent} total scattering matrix ${\mathbb S}^\phi$, leading to the effective Y-junction in the right 
hand side of fig.~\ref{fig.1}. It is worth mentioning that the {\it critical} (scale invariant) conductance properties 
of Y-junctions and their phase diagram have been previously investigated by different methods 
like bosonization \cite{nfll-99} -\cite{Bellazzini:2010gs}, 
renormalization group and scattering techniques \cite{lrs-02} -\cite{aw-11} and 
conformal field theory \cite{coa-03, rhfoca-11}. 
We concentrate below on some {\it off-critical} aspects, establishing first the exact 
form of $\S^\phi$. We discuss afterwards both the conductance and the noise at finite temperature and 
discover some new features, related to the finite size of the ring and the non-trivial magnetic flux.  

\begin{figure}[ht]
\begin{center}
\begin{picture}(220,100)(-60,10) 
\includegraphics[scale=0.8]{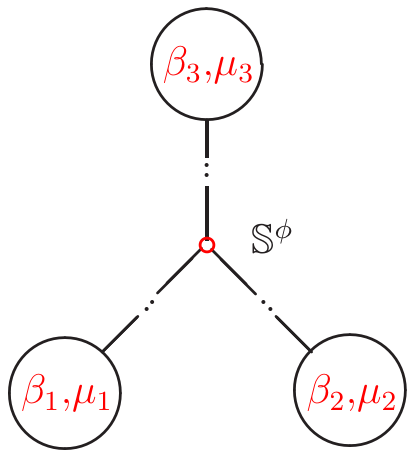}
\end{picture} 
\end{center}
\caption{(Color online) Y-junction connected at infinity to 
thermal reservoirs with temperature 
$\beta_i$ and chemical potential $\mu_i$.}
\label{fig.2}
\end{figure}

The system is driven away from equilibrium by 
attaching to the external leads thermal reservoirs at (inverse) temperatures $\beta_i$ and chemical 
potentials $\mu_i$, as shown in fig.~\ref{fig.2}. Our main goal is to study the transport properties 
and the noise of this configuration as a function of the transmission $\tc$ in the ring, the flux $\phi$, 
the temperatures $\beta_i$ and the chemical potentials $\mu_i$. 
We show that Aharonov-Bohm type oscillations with $\phi$ occur 
in both the conductance and the noise. The period of these oscillations 
equals the elementary flux quantum $\phi_0=2\pi \hbar c/e$ associated with a single charge $e$. 
We find that the pure thermal noise has a $\phi$-dependent power law behavior at small temperatures which 
interpolates between the usual linear Johnson-Nyquist behavior and a new, quadratic behavior for values of $\phi$ 
greater than a critical value $\phi_c$ which we quantify. Finally, as functions of $\mu_i$, 
the current and the shot noise show in the regime $\tc\sim 1$ an interesting plateaux structure, which is 
related to a resonant tunneling effect.  
The fundamental and essentially unique input for deriving these results is the requirement of 
self-adjointness of the Schr\"odinger Hamiltonian with magnetic flux $\phi$ on the graph in fig.~\ref{fig.1}. 

We would like to mention also that the wires, displayed in the figures of this paper, 
are planar and in most of the cases have straight line edges. 
However, the discussion below is completely general and applies 
to segments of arbitrary smooth curves in $\RR^3$ as well. 
What is essential is to have a well defined tangent vector field along the edges $E_i$, $i=1,2,3$, 
in order to define the projection $A_x(x,i)$ of the ambient field $\bA(\bx)$ on the graph.

\section{Bulk dynamics and local vertex interactions} 

The dynamics in each edge is defined by the Schr\"odinger equation 
(we adopt the natural units $c=\hbar =1$)   
\begin{equation}
\left [i \partial_t + \frac{1}{2m} \left (\partial_x - \ri e A_x(x,i)\right )^2\right ]\psi (t,x,i) = 0 \, , 
\label{eqm1}
\end{equation}
where $(x,i)$ are local coordinates on $E_i$. Besides the bulk dynamics, we have to introduce 
also the interaction at the vertices $V_j$, which represents a subtle point. 
Some recent developments \cite{ks-00, H1} in the spectral theory of operators on graphs  
have shown that these two ingredients are not independent \cite{Bellazzini:2006jb, Bellazzini:2008mn}, 
{\it if one requires unitary time-evolution of the system}. 
The reason is that the time evolution in the bulk is described by a {\it Hermitian} Hamiltonian, which 
becomes {\it self-adjoint} only by imposing special boundary conditions 
at the vertices. These conditions generate {\it particular} point-like interactions, which are 
described by specific (and not arbitrary) scattering matrices $S_j$, associated with 
each vertex $V_j$ of the graph. Let us illustrate the phenomenon using for simplicity the bulk 
Hamiltonian $-\partial_x^2$ corresponding to (\ref{eqm1}) with 
$e=0$. Assume that the vertex $V$ with local coordinate $x=0$ is the origin of $n$ edges $E_i$. 
The most general boundary condition ensuring that $-\partial_x^2$ has a self-adjoint 
extension at $x=0$ is \cite{ks-00, H1}  
\begin{equation} 
\sum_{j=1}^{n} \left [\lambda ({\mathbb I}- U)_{ij}\, 
\psi (t,0,j) -i ({\mathbb I}+ U)_{ij}
(\partial_x\psi ) (t,0,j)\right ] = 0\, , 
\label{bc1} 
\end{equation} 
where $U$ is an arbitrary $n\times n$ unitary matrix and $\lambda$ 
is a real parameter with the dimension of mass. $U=\II$ and $U=-\II$ 
generalize to a vertex with multiple edges the familiar Neumann and 
Dirichlet boundary conditions on the half line. The point-like interaction, induced by  
(\ref{bc1}), generates \cite{ks-00, H1} 
\begin{equation} 
S(k) = -\frac{[\lambda ({\mathbb I} - U) - 
k({\mathbb I}+ U )]}{[\lambda ({\mathbb I} - U) 
+ k({\mathbb I}+ U)]}\, , \qquad k\in \RR\, ,
\label{S1}
\end{equation} 
which defines a family of very special unitary momentum-dependent scattering matrices 
parametrized by $U\in U(n)$. $S(k)$ is a meromorphic function with simple poles, all of which located on 
the imaginary axis and different from 0. It turns out  \cite{Bellazzini:2009nk} 
that $S(k)$ preserves time reversal invariance if and only if $S(k)$ is 
{\it symmetric}. We will use this information when discussing below 
the breaking of time reversal symmetry caused by the magnetic flux. 

The critical (scale-invariant) points $S^c$ in the family (\ref{S1})  
capture the universal features of the local vertex interactions and play therefore a distinguished role. 
Requiring time-reversal invariance, $S^c$ are 
given \cite {Bellazzini:2009nk} by $k$-independent symmetric matrices 
belonging to the orthogonal group $O(n)$. Let us consider the case $n=3$, 
relevant for the Y-junction in fig. \ref{fig.1}, and let us 
assume that the internal edges of each local junction are equivalent 
as far as transmission and reflection are concerned. 
Labeling these edges by the indices 2 and 3, the matrix elements of $S^c$ 
must be invariant under the exchange $2\leftrightarrow 3$. These requirements fully 
determine the two one-parameter families in $O(3)$ 
\begin{equation}
S^c_\pm (\tc) =\pm \left(\begin{array}{ccc}
1-2\tc &\sqrt{2 \tc(1-\tc)}&\sqrt{2 \tc(1-\tc)}\\
\sqrt{2 \tc(1-\tc)}&\tc-1&\tc\\
\sqrt{2 \tc(1-\tc)}&\tc&\tc-1
\end{array}\right)\, , 
\label{A6}
\end{equation} 
where $\tc \in [0,1]$ is the {\it transmission coefficient} controlling the local tunneling between the 
edges 2 and 3 of the junction. Since ${\rm det}(S^c_\pm) = \pm1$, the matrices (\ref{A6}) 
belong to the two disconnected components of $O(3)$. 
The matrix $S^c_-(\tc=1/2)$ has been introduced in Ref. \cite{gia-84}. 
In \cite{biy-84} the matrices $S^c_\pm(\tc)$ have been considered 
for generic $\tc\in [0,1]$. We argued above that 
$S^c_\pm(\tc)$ are critical points in the set of all scattering matrices ensuring 
the self-adjointness of the Schr\"odinger Hamiltonian on the graph in fig. \ref{fig.1}.   

The above considerations can be extended to 
the case $e\not=0$, performing the 
shift  $\partial_x \longmapsto \partial_x -\ri e A_x(x,i)$ in eq. (\ref{bc1}). Introducing 
the magnetic flux 
\begin{equation}
\phi = \oint_C {\bf A}({\bf x})\cdot \rd {\bf \ell}\, , 
\label{flux}
\end{equation} 
where $C$ is the ring in fig.~\ref{fig.1}, the shift generates  
$e\phi$-dependent phases that charges pick up traveling along the edges. 
These phases are transferred \cite{Mintchev:2011mx, CMR} by a kind 
of ``gauge" transformation to the matrix $U$ and therefore to $S(k)$. 
We set for simplicity $e=1$ in the rest of the paper.

\section{The total scattering matrix $\S^\phi$} 

The problem now is to reconstruct the 
total scattering matrix $\S^\phi$ in fig.~\ref{fig.1} from the local ones. Several 
equivalent schemes \cite{KS}-\cite{Caudrelier:2009ay} exist for facing this problem. 
We follow below the approach of \cite{Caudrelier:2009ay}, which 
adapts better to the case with ambient magnetic field and provides explicit expressions. 
Since the form of $\S^\phi$ for a generic ring junction with general $S_i(k)$ is quite complicated,  
we simplify the considerations by focusing on the case of {\it identical}  
local scattering matrices and {\it equidistant} 
vertices, separated by a distance $d$ along the ring. In this case the system is invariant under 
cyclic permutations, implying that $\S^\phi$ is a {\it circulant} matrix, i.e. 
\begin{eqnarray}
\S_\pm^\phi(k)=\left(\begin{array}{ccc}
\sigma_1^\pm(k,\phi)&\sigma_2^\pm(k,\phi)&\sigma_3^\pm(k,\phi)\\
\sigma_3^\pm(k,\phi)&\sigma_1^\pm(k,\phi)&\sigma_2^\pm(k,\phi)\\
\sigma_2^\pm(k,\phi)&\sigma_3^\pm(k,\phi)&\sigma_1^\pm(k,\phi)
\end{array}\right)\, . 
\label{CM}
\end{eqnarray}
{}For analyzing the universal features of the Y-junction, it is enough to concentrate 
on the critical local $S$-matrices given by eq. (\ref{A6}). In this case one has \cite{CMR}
\begin{equation}
\sigma_j^\pm(k)=
\frac{1}{3}\sum_{\ell=1}^3\e^{{\ri \frac{2\pi}{3}}{(1-\ell)(j-1)}}\lambda_\pm \left(k,\frac{\phi+2(\ell-1)\pi}{3}\right)\, ,
\label{LS2}
\end{equation}
with 
\begin{equation}
\lambda_\pm (k, \theta )=
\mp \frac{\tc(\cos \theta \mp \cos kd) \pm \ri (\tc-1)\sin kd}{\tc(\cos \theta  \mp \cos kd) \mp \ri(\tc-1)\sin kd}\, .
\label{LS3}
\end{equation} 

Equations (\ref{CM}, \ref{LS2}, \ref{LS3}) represent a fundamental point of our investigation 
and determine the following set of total scattering matrices 
$\{\S^\phi_\pm (k; d, \tc)\, :\, d\geq 0,\, \tc\in [0,1]\}$. 
Since  
\begin{equation} 
\S_\pm^\phi (k; 0, \tc) = \mp \II \, ,
\label{exc1}
\end{equation}
which describe three {\it disconnected} edges, we take in the rest $d\not=0$. 
Moreover, since  
\begin{equation} 
\S^\phi_- (k; d, \tc) = - \S^{(\phi+\pi)}_+ (k; d, \tc)\, , 
\label{RS}
\end{equation}
without loss of generality we concentrate in what follows on $\S^\phi_+$, 
omitting for simplicity the index $+$. Observing that 
\begin{equation}
\sigma_2 (k,\phi) \not= \sigma_3(k,\phi) \, , \qquad {\rm for}\quad  \phi \not= 3 n \pi \, , \quad n \in \ZZ\, , 
\label{s23}
\end{equation} 
we conclude that time reversal invariance is broken ($\S^\phi$ is not symmetric), 
except for the fluxes $\phi = 3 n \pi$. 

\begin{figure}[ht]
\begin{center}
\includegraphics[scale=0.50]{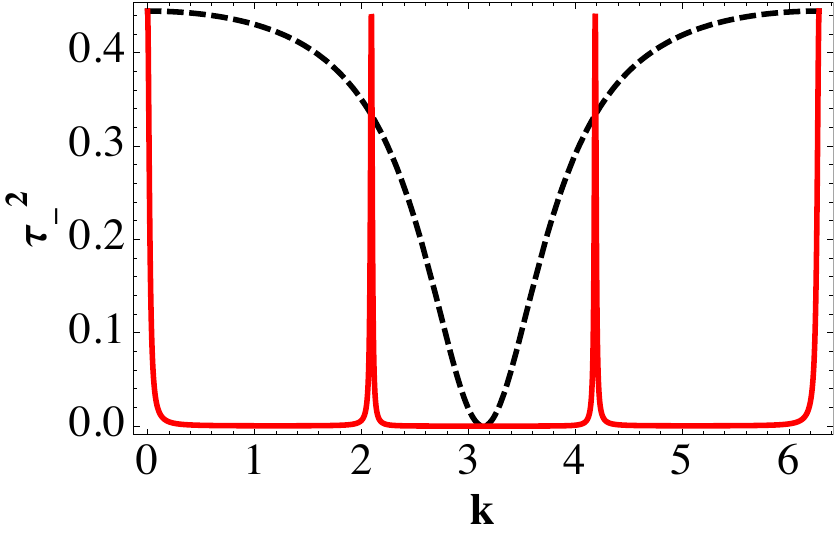}
\includegraphics[scale=0.50]{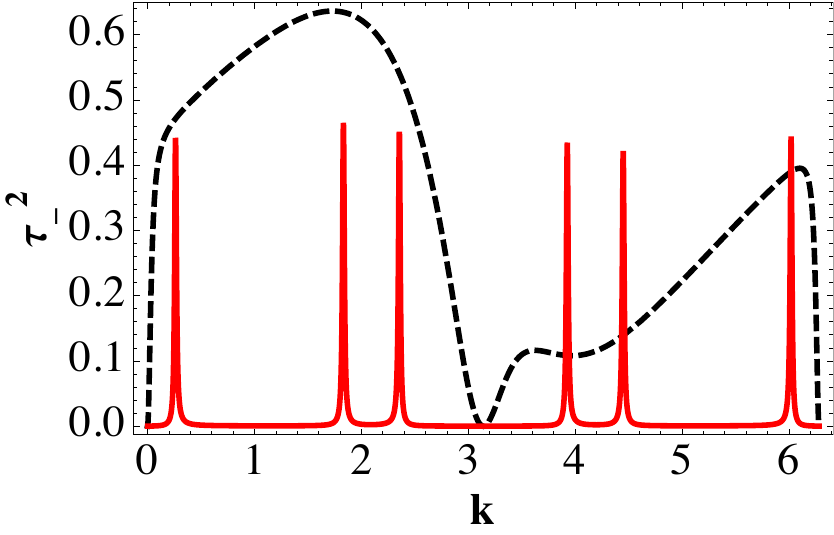}
\end{center}
\caption{(Color online) $\tau_-^2 (k, \phi, d=1)$ for $\tc=0.5$ (dashed black line) 
and $\tc=0.99$ (continuous red line) for $\phi=0$ (left) 
and $\phi = \pi/4$ (right).} 
\label{fig.3}
\end{figure}

We focus at this point on the {\it transmission amplitudes}
\begin{equation}
\tau_+(k,\phi)\equiv|\sigma_2(k,\phi)|\, ,\qquad \tau_-(k,\phi) = |\sigma_3(k,\phi)| \, ,  
\label{s23t}
\end{equation} 
and the {\it reflection amplitude} 
\begin{equation}
\varrho (k,\phi) \equiv |\sigma_1(k,\phi)|\, , 
\label{s1r}
\end{equation} 
which satisfy (due to unitarity) the expected relation 
\begin{equation}
\varrho^2(k,\phi) + \tau_+^2(k,\phi)+\tau_-^2(k,\phi) = 1\, , 
\label{rtprob}
\end{equation}
among probabilities. 
One can verify that both $\tau_\pm$ and $\varrho$ are periodic in $\phi$ with 
period $\phi_0 = 2\pi$. The period $\phi_0$ has 
a deep physical meaning. In fact, recalling our convention $c=\hbar =e=1$, $\phi_0$ equals  
precisely the elementary flux quantum $2\pi \hbar c/e$ associated with a single charge $e$ and 
appearing \cite{ASh-87} in the context of Aharonov-Bohm type oscillations in 
non simply connected mesoscopic systems. Note also that 
\bea
\label{parity}
\varrho(k,\phi)=\varrho(k,-\phi)~~,~~\tau_-(k,\phi)=\tau_+(k,-\phi)
\eea
Hence, we 
can restrict $\phi$ to $[0,2\pi]$ and even to $[0,\pi]$ when dealing with quantities involving only $\varrho$.
We emphasize that both $\varrho$ and $\tau_\pm$ are $k$-dependent, in spite of 
the fact that the local scattering matrices (\ref{A6}) are constant. This dependence 
is a direct consequence of the finite size of the ring. In fact, the momentum $k$ enters 
$\varrho$ and $\tau_\pm$ only through the dimensionless combination $kd$. It follows from (\ref{LS3}) 
that $\varrho$ and $\tau_\pm$ are $2\pi/d$-periodic in $k$, the shape of the oscillations being 
strongly influenced by the transmission $\tc$ and the flux $\phi$. The behavior of the probability 
$\tau^2_-$ for $d=1$, shown in fig.~ \ref{fig.3}, confirms this statement. 
The dashed (black) and the continuous (red) lines describe the oscillations for 
$\phi = 0$ (left) and $\phi=\pi/4$ (right) for $\tc=0.5$ and $\tc=0.99$ respectively. 
As already observed, for $\tc\sim 1$ 
the external edges are almost isolated. Accordingly, in this regime, one expects very 
small transmission amplitudes. This is indeed the case with the exception of 
the momenta $k=\frac{\pm \phi + 2\pi n}{3d}$, characterized by the appearance 
of sharp peaks with maximum close to $4/9$ and corrections of order $1-\tc$. 
For $\phi\neq n\pi$, there are six of them in each $k$ interval of length $2\pi/d$ and 
only three if $\phi=n\pi$. fig.~ \ref{fig.3} illustrates the 
phenomenon for $\phi=0$ (left) and $\phi = \pi/4$ (right). 
Because of (\ref{rtprob}), the behavior of the reflection amplitude $\varrho$ 
is complementary. 

We stress that at $\tc=1$ the amplitudes $\tau_\pm$ become actually {\it discontinuous} in $k$. 
One has indeed 
\begin{equation}
\lim_{\tc\to 1} \tau_\pm (k,\phi) = 
\begin{cases}
\frac{2}{3}\, , & \quad k = \frac{\pm \phi + 2\pi n}{3d}\, ,\qquad n \in \ZZ\, , \\
0\, , & \quad k \not= \frac{\pm \phi + 2\pi n}{3d}\, . 
\end{cases}
\label{disc1}
\end{equation}
This special $k$-dependence of $\tau_\pm$ for $\tc\sim 1$ is at the origin of the resonant tunneling effect on the current and 
the shot noise discussed below. 
Similar phenomena show up 
in the other limit $\tc\sim 0$ of almost disconnected external edges. 

Another type of discontinuities of $\tau_\pm$, which involves the magnetic flux and also 
deserves attention, is described by 
\begin{equation}
\lim_{k\to \frac{2l\pi}{d}} \tau_\pm (k, \phi) = 
\begin{cases}
\frac{2}{3}\, , & \quad \phi = n\phi_0\, ,\qquad l,\, n \in \ZZ\, , \\
0\, , & \quad \phi \not= n\phi_0\, . 
\end{cases}
\label{disc2}
\end{equation} 
This behavior is at the origin of the effect on 
the pure thermal noise at small temperatures discussed below. 

\section{Currents, conductance and noise} 

To the end of the paper we study the non-equilibrium transport properties of the Y-junction in fig.~ \ref{fig.2} 
with thermal reservoirs at inverse temperatures $\beta_i$ and chemical potentials 
\begin{equation}
\mu_i = k_F - V_i \, , \qquad i=1,2,3\, , 
\label{cp}
\end{equation}
where $k_F$ defines the Fermi energy and $V_i$ is the external voltage applied to the edge $E_i$. 
In what follows we keep $k_F$ fixed, varying eventually the gate voltages $V_i$.  
The system is away from equilibrium if $\S^\phi $ admits at 
least one non-trivial transmission coefficient among edges with different $\beta_i$ and/or $\mu_i$. 
The corresponding non-equilibrium dynamics is implemented by a 
{\it steady state} $\Omega_{\beta,\mu}$, characterized by non-vanishing time-independent charge 
and heat currents circulating along the leads. The construction \cite{Mintchev:2011mx} of 
$\Omega_{\beta,\mu}$ involves the scattering matrix 
$\S^\phi$ and fully takes into account both the minimal coupling with the external 
magnetic field and the vertex interactions. We denote in what follows the expectation values 
in the state $\Omega_{\beta,\mu}$ by $\langle \cdots \rangle_{\beta,\mu}$ and stress  
that the current correlation functions 
$\langle j_x(t,x,i) \rangle_{\beta,\mu}$ and $\langle j_x(t_1,x_1,i_1) j_x(t_2,x_2,i_2)\rangle_{\beta,\mu}$ 
used below are \textit{exact}. No approximations, like linear response theory, are adopted. 

{}For the one-point function one finds the Landauer-B\"uttiker \cite{la-57, bu-86} expression  
\begin{equation}
J_i \equiv \langle j_x(t,x,i) \rangle_{\beta, \mu}  
= \int_0^\infty \frac{\rd k}{2\pi} \frac{k}{m}\sum_{j=1}^3  \left [\delta_{ij} -  
|\S^\phi_{ij}(k)|^2\right ] d_j(k) 
\label{LB}
\end{equation} 
where 
\begin{equation} 
d_i (k) = \frac{\e^{-\beta_i [\omega (k) - \mu_i]}}{1+ \e^{-\beta_i [\omega (k) - \mu_i]}} \, , 
\quad \omega (k) = \frac{k^2}{2m}\, , 
\label{fbd1} 
\end{equation}
is the familiar Fermi distribution. The periodicity of $|\sigma_i(k)|$ in $\phi$ implies that $J_i$ oscillate 
with period $\phi_0$. The unitarity of $\S^\phi$ leads to the $\tc$-independent bound 
\begin{equation}
|J_i| \leq \frac{1}{2 \pi \beta_i}\log\left [1+ \e^{(k_F-V_i)\beta_i}\right ] \, , \quad 
\label{est1}
\end{equation}
on the amplitude of the oscillations. Introducing the variable $\xi = k^2/2m$ and 
using that the Fermi distribution (\ref{fbd1}) approaches 
the Heaviside step function $\theta(\mu_i-\xi)$ in the 
zero temperature limit $\beta_i \to \infty$, one obtains from (\ref{LB}) 
\begin{equation}
J_i = 
\theta(\mu_i) \frac{\mu_i}{2\pi} - 
\sum_{j=1}^3 \theta (\mu_j) \int_0^{\mu_j} \frac{\rd \xi}{2\pi} 
\left |\S^\phi_{ij}\left (\sqrt {2m\xi}\right )\right |^2.  
\label{ztc}
\end{equation} 
We will compare below (\ref{ztc}) to the shot noise. 

\begin{figure}[ht]
\begin{center}
\includegraphics[scale=0.50]{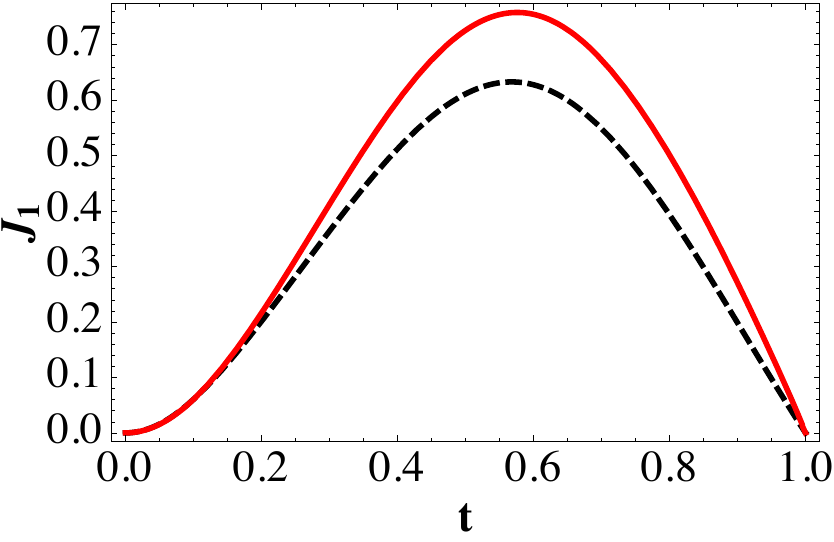}
\includegraphics[scale=0.50]{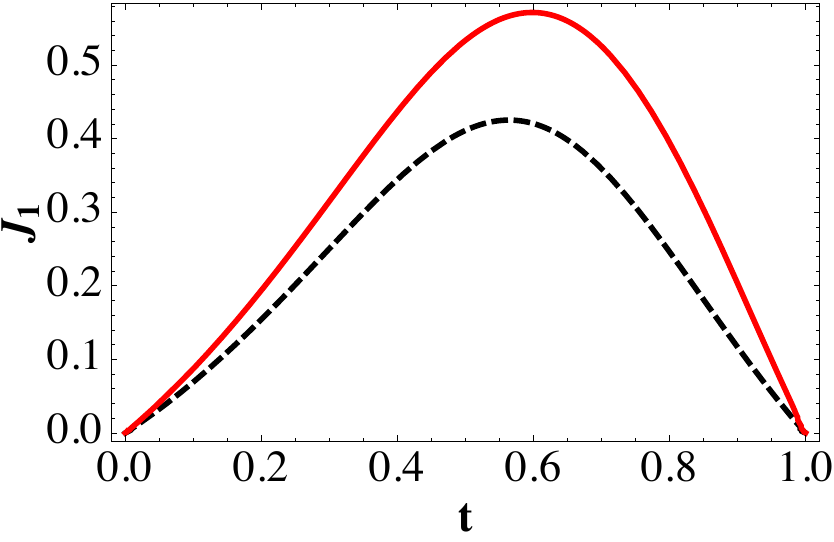}
\includegraphics[scale=0.50]{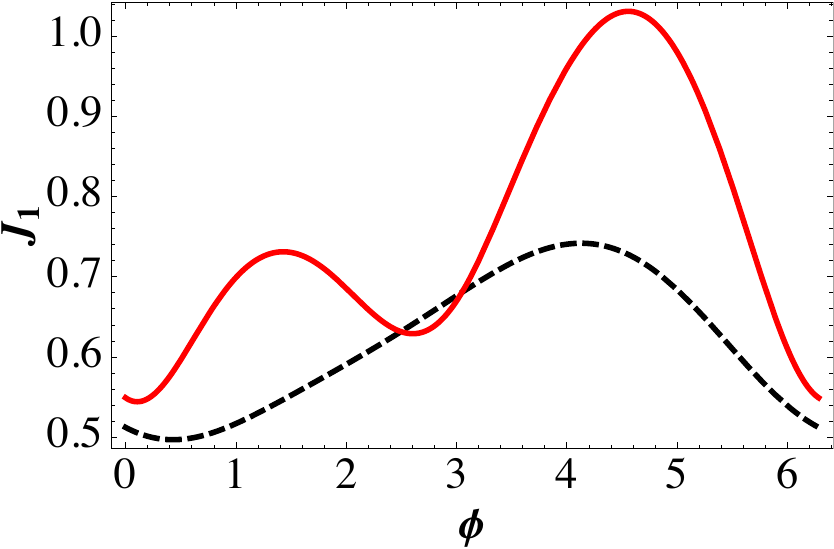}
\includegraphics[scale=0.50]{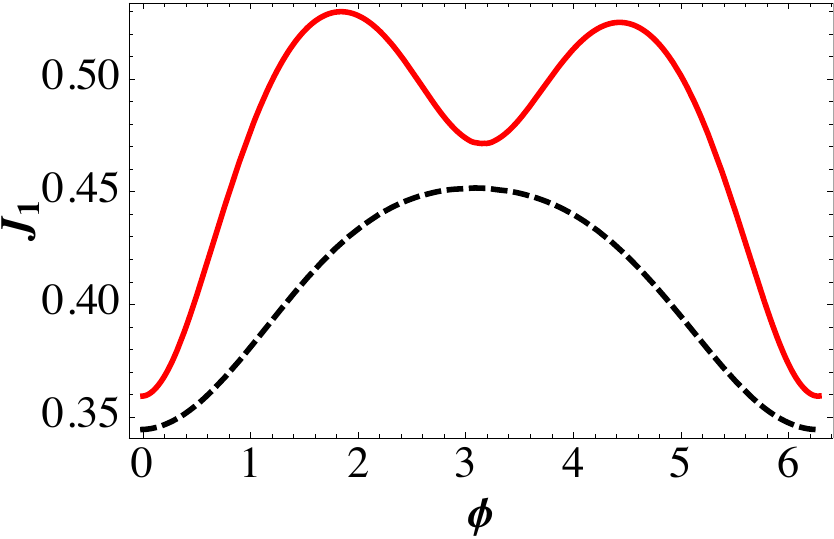}
\end{center}
\caption{(Color online) $J_1(\phi,\tc)$  
for fixed $\phi$ (first line) and fixed $\tc$ (second line) 
at $\beta_i=10$ (left) and $\beta_i=0.1$ (right).} 
\label{fig.4}
\end{figure}

In order to get a more precise idea on the dependence of the current $J_i$ on $\phi$, the 
transmission $\tc$ and the temperature $\beta$, we concentrate on (\ref{LB}). 
The $k$-integration can not be performed in a closed analytic form, but being well defined, the integral 
can be computed numerically. The plots in fig. \ref{fig.4} illustrate the result for 
$d=1$, $m=1/2$, $k_F=3$, $V_1 =-V_2=-5$ and $V_3=0$. 
The first line displays $J_1$ as a function of $\tc$ for 
$\phi=0$ (dashed line) and $\phi = 2\pi/3$ (continuous line) at 
$\beta_1 = \beta_2 =\beta_3 =10$ (left) and $\beta_1 = \beta_2 =\beta_3 =0.1$ (right). 
As expected, the current vanishes at $\tc=0,1$, when the external edges 
are isolated from each other. We see also that the 
position of the maximum of $J_1$ depends on the flux and the temperature. 
In the second line of fig. \ref{fig.4} we report $J_1$ as a function of $\phi$ at 
$\tc=0.4$ (dashed) and $\tc=0.7$ (continuous), which show the expected oscillation in $\phi$ 
with period $\phi_0$. The continuous (red) lines illustrate the impact of the higher 
harmonics $n \phi_0$, which become relevant for $\tc >0.5$. 

\begin{figure}[ht]
\begin{center}
\includegraphics[scale=0.5]{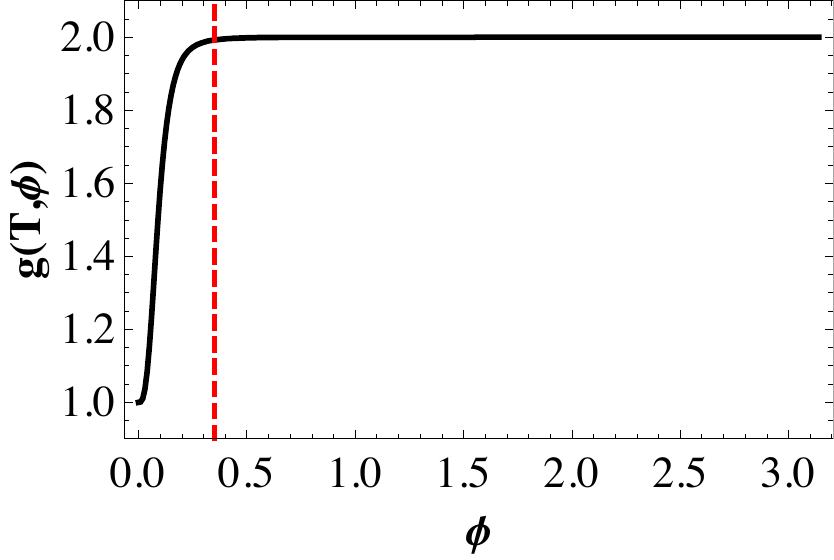}
\includegraphics[scale=0.5]{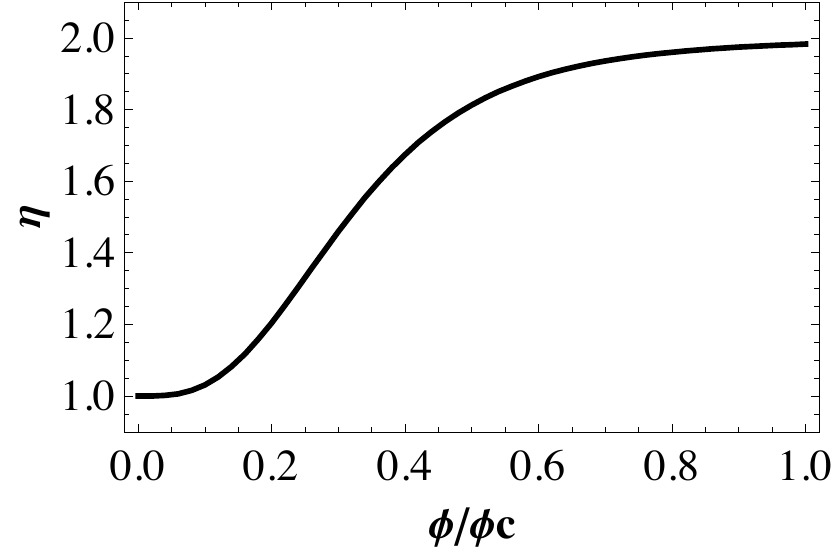}
\end{center}
\caption{(Color online) Left: full behavior of $g(T,\phi)$ (for $T=10^{-5}$ and other parameters as before). Dashed line shows $\phi_c$. Right: universal profile $\eta$
as a function of the rescaled flux $\phi/\phi_c$ (zoom of the interpolation region).} 
\label{fig.5}
\end{figure}

The {\it zero frequency noise power} is defined as usual\cite{bb-00} by 
\begin{equation}
P_{ij} = \lim_{\nu \to 0^+} \int_{-\infty}^\infty \rd t\,  \e^{i \nu t} \, 
\langle j_x(t,x_1,i) j_x(0,x_2,j) \rangle_{\beta, \mu}^{\rm conn}\, , 
\label{N1}
\end{equation}
where 
$\langle j_x(t_1,x_1,i) j_x(t_2,x_2,j) \rangle_{\beta, \mu}^{\rm conn}$ 
is the {\it connected} current-current correlation function in the state $\Omega_{\beta,\mu}$. 
It turns out\cite{Mintchev:2011mx} that $P_{ij}$ is $x_{1,2}$-independent and is given by 
\begin{eqnarray}
P_{ij} = \frac{1}{m} \int^{\infty}_0 \frac{\rd k}{2 \pi} k 
\Bigl \{ \delta_{ij} D_{ii}(k)  - |\S^\phi_{ij}(k)|^2 D_{jj}(k) \qquad 
\nonumber \\
- |\S^\phi_{ji}(k)|^2 D_{ii}(k) + \sum_{l,m=1}^3 {\cal F}_{ijlm}(k)[D_{lm}(k) + D_{ml}(k)]\Bigr \}\, , 
\label{N2}
\end{eqnarray}
where 
\begin{equation}
{\cal F}_{ijlm}(k) = \frac{1}{2}\S^\phi_{il}(k)\overline{\S}_{jl}^\phi(k) 
\S^\phi_{jm}(k)\overline{\S}^\phi_{im}(k) 
\end{equation}
and $D_{ij}(k) \equiv d_i(k)[1-d_j(k)]$.
$P_{ij}$ is a symmetric matrix. To study the thermal noise, we assume $\mu_i=\mu$ and 
$\beta_i=\beta$. Using the Kirchhoff rule $\sum_{i=1}^3 P_{ij} =0$, we get 
the circulant matrix
\begin{equation} 
P_{ij} = P \left (3\,\delta_{ij}-1\right )/2
\label{pnew}
\end{equation}
where
\begin{equation}
P\equiv P(T,\phi) =\frac{2}{m} \int^{\infty}_0 
\frac{\rd k}{2 \pi}\, k \,D(k)\, \left [\tau_-^2(k,\phi) +\tau_+^2(k,\phi)\right ] 
\label{N3}
\end{equation}
with $D(k) \equiv d(k)[1-d(k)]$ and $T=1/\beta$.
Like the conductance, $P$ oscillates in $\phi$ with period $\phi_0$. From (\ref{parity}), $P$ is an even function of $\phi$ 
so it is enough to study it on $[0,\pi]$. The bound on the amplitude, 
following from unitarity is now  
\begin{equation}
0\leq P \leq \frac{2}{m} \int^{\infty}_0 \frac{\rd k}{2 \pi}\, k \, D(k) = 
\frac{1}{\pi \beta \left (1+\e^{-\beta \mu}\right )}\, . 
\label{est2}
\end{equation} 
Define
\bea
g(T,\phi)=\frac{\partial \ln P(T,\phi)}{\partial \ln T}\,.
\eea
The numerical study confirms that $g=1 $ at large temperatures $T \to \infty$, 
independently of the flux $\phi$. In this regime 
one recovers therefore the well-known Johnson-Nyquist behavior $P\sim T$. The situation changes 
drastically as $T \to 0$. For $\mu=0$, the pure thermal noise has the following power law type behavior 
as $T\to 0$
\begin{equation}
g(T,\phi )=
\begin{cases}
1\, , & \quad \phi = 0\, , \\
\eta(\phi)\,,&\quad\,0\le \phi\le \phi_c\,,\\
2\, , & \quad \phi_c\le\phi\le \pi\,,
\end{cases}
\label{N5}
\end{equation} 
where the critical value $\phi_c$ scales like $(mT)^{\frac{1}{4}}$ and $\eta$ is a 
universal profile, independent of $T$, interpolating between the linear ($g=1$ at $\phi=0$) and quadratic ($g=2$ for $\phi>\phi_c$) behavior of $P$.
This is shown in fig. \ref{fig.5}.
These results can also be derived analytically. The details will be reported elsewhere \cite{CMR}. Note that this behavior is also 
valid for nonzero $\mu$ as long as $\mu/T<<1$.

Therefore, we conclude that a nonzero magnetic flux 
in the Y-junction implies a significant 
modification of the Johnson-Nyquist law at small temperature. 
This new feature provides an interesting signature of a physical effect that hopefully can 
be observed experimentally. 

Let us investigate finally the shot noise following from eq.(\ref{N2}). For this purpose 
we set $\beta_i=\beta$ and take the $\beta\to\infty$ limit, keeping $\mu_i >0$ arbitrary. 
Adopting the variable $\xi = k^2/2m$ one gets 
\begin{equation}
P_{ij} = \sum_{l\not=m=1}^3\varepsilon(\mu_l-\mu_m)\int_{\mu_m}^{\mu_l} 
\frac{\rd \xi}{2\pi}\, \cF_{ijlm}\left (\sqrt{2m\xi}\right )\, ,
\label{SN1}
\end{equation}
$\varepsilon $ being the sign function. Note that since $P_{ij}$ is symmetric 
and satisfies the Kirchhoff rule, we only need to compute the diagonal elements 
$P_{ii}$ in order to reconstruct the complete matrix. Indeed, one has 
$P_{ij} =\frac{1}{2}(P_{kk}-P_{ii}-P_{jj})$ for mutually distinct $i,j,k$ and the diagonal elements read 
\begin{equation}
P_{ii} = 2 \sum_{l<m=1}^3\varepsilon(\mu_l-\mu_m)\int_{\mu_m}^{\mu_l} \frac{\rd \xi}{2\pi}\, 
\cF_{iilm}\left (\sqrt{2m\xi}\right ) \, .  
\label{SN3}
\end{equation}
Assuming for definiteness that $\mu_1 <\mu_2 <\mu_3$, one obtains 
\begin{equation}
P_{11} = \int_{\mu_1}^{\mu_2} \frac{\rd \xi}{2\pi}\, \varrho^2(1-\varrho^2) + 
\int_{\mu_2}^{\mu_3} \frac{\rd \xi}{2\pi}\, \tau_-^2(1-\tau_-^2)\, , 
\label{SN4}
\end{equation}
\begin{equation}
P_{22} = \int_{\mu_1}^{\mu_2} \frac{\rd \xi}{2\pi}\, \tau_-^2(1-\tau_-^2)+
\int_{\mu_2}^{\mu_3} \frac{\rd \xi}{2\pi}\, \tau_+^2(1-\tau_+^2)\, , 
\label{SN5}
\end{equation}
\begin{equation}
P_{33} = \int_{\mu_1}^{\mu_2} \frac{\rd \xi}{2\pi}\, \tau_+^2(1-\tau_+^2) + 
\int_{\mu_2}^{\mu_3} \frac{\rd \xi}{2\pi}\, \varrho^2(1-\varrho^2)\,,
\label{SN6}
\end{equation}
where $\varrho$ and $\tau_\pm$ are computed at $k=\sqrt{2m\xi}$. Compared to the 
pure thermal noise (\ref{pnew},\ref{N3}), the shot noise involves the fourth order powers of 
$\varrho$ and $\tau_\pm$ as well. Their dependence on $\phi$ implies that 
$P_{ii}$ oscillate with period $\phi_0$. The amplitude is subject to the obvious 
unitarity bound 
\begin{equation}
0\leq P_{ii} \leq \mu_3 -\mu_1 \, . 
\label{SN7}
\end{equation}

We study finally the behavior of the shot noise as a function of the 
chemical potentials $\mu_i$, or equivalently, the voltages $V_i$ in (\ref{cp}). 
It is instructive to do this, comparing $P_{ii}$ with the 
zero-temperature steady current $J_i$ given by (\ref{ztc}), and the transmission 
amplitude $\tau_\pm$. For this purpose we fix $m=1/2$, $\mu_1=\mu_2 = d= 1$ and vary $\mu_3$. 
In this regime 
\begin{equation}
J_1 (\phi)= - \int_1^{\mu_3}\frac{\rd \xi}{2\pi}\, \tau_-^2\, , \quad 
P_{11}(\phi) = \int_{1}^{\mu_3} \frac{\rd \xi}{2\pi}\, \tau_-^2(1-\tau_-^2)\, .  
\label{SN8}
\end{equation}

\begin{figure}[ht]
\begin{center}
\includegraphics[scale=0.5]{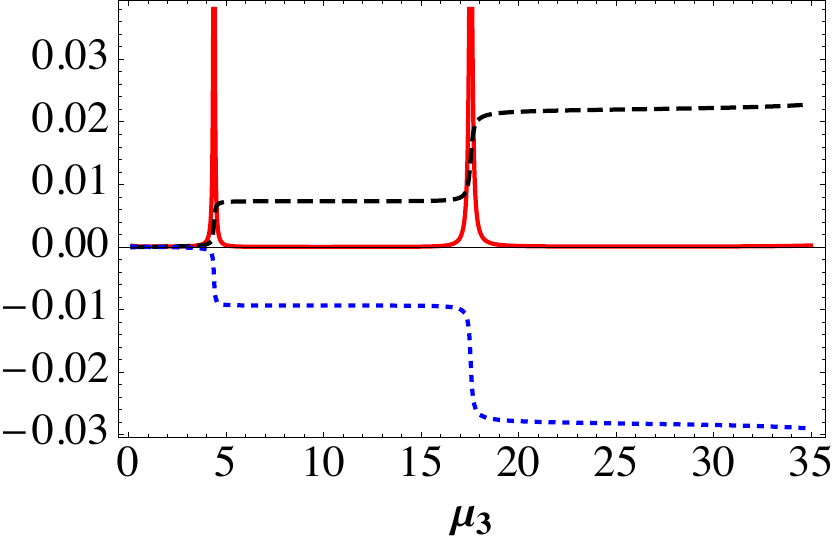}
\includegraphics[scale=0.5]{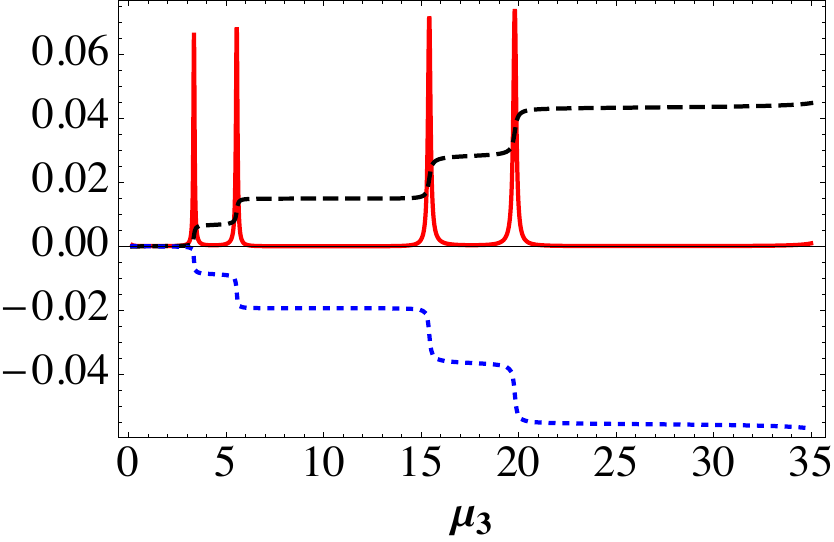}
\end{center}
\caption{(Color online) Plots of $P_{11}$ (black dashed), $J_1$ 
(blue dotted)  and $\tau_-^2/2\pi$ (red continuous) at $\tc=0.99$ 
for $\phi=0$ (left) and $\phi =\pi/4$ (right).} 
\label{fig.6}
\end{figure}
An interesting resonant tunneling effect, depending on $\phi$, 
is observed for $\tc\sim 1$. This corresponds to the situation where the external 
edges are weakly coupled to the ring. The peaks in the transmission amplitudes $\tau_\pm$, 
shown in fig. \ref{fig.3}, can be interpreted as resonances 
corresponding to eigenstates of the ring. A similar situation was discussed in \cite{biy-84, rs-04} in the case of the ring with 
two external edges and the same physical interpretation holds here. As the voltage is increased, these resonances generate plateaux in 
the shot noise $P_{ii}$ and the current $J_i$. This fact is 
illustrated in fig. \ref{fig.6}, where we plotted $\tau^2_- (\sqrt {\mu_3})/2\pi$ (continuous red curve), 
$P_{11}(\mu_3)$ (dashed black curve) and $J_1(\mu_3)$ (dotted blue curve). 
Switching on the magnetic field changes the location of the peaks and hence the location of the jumps from 
one plateau to the next. 

\section{Conclusions} 
The transport properties of fermions in a Y-junction with a finite size ring, connected to thermal reservoirs 
and crossed by magnetic flux $\phi$, have been investigated. The bulk dynamics is 
described by the Schr\"odinger equation with the minimal coupling to an ambient electromagnetic field. 
At the vertices where the external leads are attached to the ring, the most general scale invariant 
local interactions, compatible with a unitary time evolution, are considered.  
The exact expression for the total scattering matrix $\S^\phi$ of the system  
is fundamental for our investigation. The non-equilibrium dynamics, 
generated by the contact to the heat baths, is captured by steady states $\Omega_{\beta, \mu}$ 
incorporating $\S^\phi$. It is essential that our framework 
does not rely on conformal symmetry, thus allowing us to investigate directly a finite size ring. 
The conductance and the noise power are extracted from the 
current correlation functions in the state $\Omega_{\beta, \mu}$. 
We find a resonant tunneling effect when the ring is weakly 
coupled to the external leads. Another interesting phenomenon 
concerns the influence of the magnetic flux on the noise (and conductance). 
For $\phi \not= 0$, we found a drastic departure from the linear 
Johnson-Nyquist law for small temperatures. Let us mention in this respect that the same analysis applies to a Dirac Y-junction 
which shows an interpolation between linear and \textit{cubic} (instead of quadratic) power law behavior, 
the difference being a consequence of the linear dispersion relation of the Dirac equation \cite{CMR}. 

All the physical information about the Y-junction has been extracted in our discussion 
from the one-point and two-point current correlation functions. It will be interesting to 
extend the above analysis to the higher correlators, thus investigating the 
effect of the magnetic flux on the full counting statistics \cite{fcs}, which provides 
further details about the system.


\begin{thebibliography}{0} 


\bibitem{nfll-99} 
C. Nayak, M. P. A. Fisher, A.~W.~W.~Ludwig and H.~H.~Lin, 
Phys.\ Rev.\  B {\bf 59} (1999) 15694.

\bibitem{sdm-01} I. Safi, P. Devillard, and T. Martin, 
Phys. Rev. Lett. {\bf 86} (2001) 4628. 

\bibitem{rs-04} S. Rao and D. Sen, Phys. Rev. B {\bf 70} (2004) 195115. 

\bibitem{Bellazzini:2006kh} 
B.~Bellazzini, M.~Mintchev and P.~Sorba,
J.\ Phys.\ A  {\bf 40} (2007) 2485.

\bibitem{hc-08}
C.-Y. Hou and C. Chamon, Phys. Rev. B {\bf 77} (2008) 155422.

\bibitem{dr-08} S. Das and S. Rao, Phys. Rev. B {\bf 78} (2008) 205421. 

\bibitem{Bellazzini:2008fu} 
B.~Bellazzini, P.~Calabrese and M.~Mintchev, 
Phys.\ Rev.\  B {\bf 79} (2009) 085122. 

\bibitem{Bellazzini:2009nk} 
B.~Bellazzini, M.~Mintchev and P.~Sorba, 
Phys.\ Rev.\  B {\bf 80} (2009) 25441.

\bibitem{Bellazzini:2010gs}
B.~Bellazzini, M.~Mintchev and P.~Sorba, 
Phys.\ Rev.\  B {\bf 82} (2010) 195113.


\bibitem{lrs-02} 
S. Lal, S. Rao, and D. Sen, Phys. Rev. B {\bf 66} (2002) 165327. 

\bibitem{emabms-05} 
X.~Barnabe-Theriault, A.~ Sedeki, V.~Meden, K.~Sch\"onhammer, 
Phys. Rev. Lett. {\bf 94} (2005) 136405.

\bibitem{drs-06} S.~Das, S.~Rao, D.~Sen, Phys. Rev. B {\bf 74} (2006) 045322. 

\bibitem{dr-09} 
S. Das, S. Rao and A. Saha, Phys. Rev. B {\bf 79} (2009) 155416. 

\bibitem{ss-11} 
A. Soori and D. Sen, Europhys. Lett. {\bf 93} (2011) 57007. 

\bibitem{Arist} 
D.~N.~Aristov, 
Phys.\ Rev.\  B {\bf 83} (2011) 115446. 

\bibitem{aw-11} 
D.~N.~Aristov and P.~W\"olfle, 
Phys.\ Rev.\  B {\bf 84} (2011) 155426. 



\bibitem{coa-03} 
M. Oshikawa,  C. Chamon, and I. Affleck,  J. Stat. Mech. (2006) P02008. 

\bibitem{rhfoca-11} 
A. Rahmani, C.-Y. Hou, A. Feiguin, M. Oshikawa, C. Chamon and I. Affleck, 
Phys. Rev. B {\bf 85}  (2012) 045120. 


\bibitem{ks-00}
V.~Kostrykin and R.~Schrader, 
Fortschr. Phys. {\bf 48} (2000) 703. 

\bibitem{H1}
M.~Harmer, 
J.\ Phys.\ A {\bf 33} (2000) 9015. 

\bibitem{Bellazzini:2006jb} 
B.~Bellazzini and M.~Mintchev, 
J.\ Phys.\ A  {\bf 39} (2006) 11101. 

\bibitem{Bellazzini:2008mn} 
B.~Bellazzini, M.~Burrello, M.~Mintchev and P.~Sorba,
Proc. Symp. Pure Math. {\bf 77} (2008) 639. 



\bibitem{KS} 
V.~Kostrykin and R.~Schrader, 
J. Math. Phys. {\bf 42} (2001) 1563. 

\bibitem{Mintchev:2007qt} 
M.~Mintchev and E.~Ragoucy, 
J.\ Phys.\ A  {\bf 40} (2007) 9515. 

\bibitem{Ragoucy:2009hf}
E.~Ragoucy, 
J.\ Phys.\ A  {\bf 42} (2009) 295205. 

\bibitem{Khachatryan:2009xg} 
S.~Khachatryan, A.~Sedrakyan and P.~Sorba, 
Nucl.\ Phys.\  B {\bf 825} (2010) 444.

\bibitem{Caudrelier:2009ay} 
V.~Caudrelier, E.~Ragoucy, 
Nucl.\ Phys.\  B {\bf 828} (2010) 515. 

\bibitem{Mintchev:2011mx}
M.~Mintchev, 
J.\ Phys.\ A  {\bf 44} (2011) 415201.

\bibitem{la-57}
R. Landauer, 
IBM J. Res. Dev. {\bf 1} (1957) 233;  
Philos. Mag. {\bf 21} (1970) 863.

\bibitem{bu-86}
M. B\"uttiker, 
Phys. Rev. Lett. {\bf 57} (1986) 1761; 
IBM J. Res. Dev. {\bf 32} (1988) 317. 


\bibitem{bb-00} 
Ya.~Blanter and M.~B\"uttiker, 
Phys. Rep. {\bf 336} (2000) 1. 
 

\bibitem{gia-84}
Y. Gefen, Y. Imry and M. Ya. Azbel, 
Phys. Rev. Lett. {\bf 52} (1984) 129.  

\bibitem{biy-84}
M. B\"uttiker, Y. Imry and M. Ya. Azbel, 
Phys.\ Rev.\  A {\bf 30} (1984) 1982. 


\bibitem{daa-94}
M. A. Davidovich and E. V. Anda, 
Phys.\ Rev.\  B {\bf 50} (1994) 15453. 

\bibitem{CMR}
V.~Caudrelier, M.~Mintchev and E.~Ragoucy, in preparation. 

\bibitem{ASh-87} 
A. G. Aronov and Yu. V. Sharvin, 
Rev. Mod. Phys. {\bf 59} (1987) 755. 

\bibitem{fcs}
L.~S.~Levitov, H.-W.~Lee and G.~B.~Lesovik, 
J.~Math.~Phys. {\bf 37} (1996) 4845.  



\end{thebibliography}
\end{document}